\documentclass[12pt,eqno,epsf]{article}
\usepackage{amsmath,amssymb,graphicx,mathtools}
\numberwithin{equation}{section}

\def\ignore#1{{}}

\newcounter{sxn}

\newcounter{axn}

\date{}

\newdimen\mybaselineskip
\renewcommand{\thefootnote}{\arabic{footnote}}

\newcommand{\beeq}{\begin{equation}}
\newcommand{\eneq}{\end{equation}}
\newcommand{\beqn}{\begin{eqnarray}}
\newcommand{\eeqn}{\end{eqnarray}}

\newcommand{\alp}{\alpha}
\newcommand{\bt}{\beta}
\newcommand{\gm}{\gamma}
\newcommand{\Gm}{\Gamma}
\newcommand{\dlt}{\delta}

\newcommand{\ep}{\epsilon}
\newcommand{\tht}{\theta}

\newcommand{\vth}{\vartheta}

\newcommand{\lmd}{\lambda}

\newcommand{\Sgm}{\Sigma}

\newcommand{\be}{\begin{equation}}
\newcommand{\ee}{\end{equation}}
\newcommand{\bea}{\begin{eqnarray}}
\newcommand{\eea}{\end{eqnarray}}
\newcommand{\eql}{\!\!\!&=\!\!\!&}

\newcommand{\sma}{\!\!\!&\simeq\!\!\!&}
\newcommand{\defa}{\!\!\!&\equiv\!\!\!&}

\newcommand{\tl}[1]{\tilde{#1}}
\newcommand{\bdm}[1]{{\mbox{\boldmath $#1$}}}

\newcommand{\der}{\partial}

\newcommand{\ie}{{i.e.}}

\newcommand{\sumint}[1]{\underset{#1}{\sum\hspace{-6mm}\int\hspace{2mm}}}

\newcommand{\vev}[1]{\langle #1 \rangle}

\newcommand{\brkt}[1]{\left( #1 \right)}
\newcommand{\brc}[1]{\left\{ #1 \right\}}
\newcommand{\sbk}[1]{\left[ #1 \right]}
\newcommand{\abs}[1]{\left| #1 \right|}

\renewcommand{\Re}{{\rm Re}\,}

\newcommand{\cA}{{\cal A}}

\newcommand{\cC}{{\cal C}}

\newcommand{\cF}{{\cal F}}
\newcommand{\cG}{{\cal G}}

\newcommand{\cL}{{\cal L}}

\newcommand{\cN}{{\cal N}}
\newcommand{\cO}{{\cal O}}

\newcommand{\cV}{{\cal V}}
\newcommand{\cU}{{\cal U}}

\newcommand{\hc}{{\rm h.c.}}

\begin{document}
\thispagestyle{empty}

\baselineskip=12pt


\begin{flushright}
KEK-TH-2228
\end{flushright}

\baselineskip=25pt plus 1pt minus 1pt

\vskip 1.5cm

\begin{center}
{\LARGE\bf KK-mode contribution to the crossover scale for the brane-induced force} 

\vspace{1.0cm}
\baselineskip=20pt plus 1pt minus 1pt

\normalsize

{\large\bf Yutaka Sakamura}${}^{1,2}\!${\def\thefootnote{\fnsymbol{footnote}}
\footnote[1]{E-mail address: sakamura@post.kek.jp}}

\vskip 1.0em

${}^1${\small\it KEK Theory Center, Institute of Particle and Nuclear Studies, 
KEK, \\ 1-1 Oho, Tsukuba, Ibaraki 305-0801, Japan} \\ \vspace{1mm}
${}^2${\small\it Department of Particles and Nuclear Physics, \\
SOKENDAI (The Graduate University for Advanced Studies), \\
1-1 Oho, Tsukuba, Ibaraki 305-0801, Japan}

\end{center}

\vskip 1.0cm
\baselineskip=20pt plus 1pt minus 1pt

\begin{abstract}
We discuss contributions of the KK modes to the crossover length scale~$r_{\rm c}$  
for the brane-induced force when the brane is given by a solitonic background field. 
We work in a 5D scalar model with a domain-wall background that mimics the DGP model. 
In spite of the infinite number of the KK modes, the crossover scale remains finite 
due to the warping effect on the ambient space of the domain wall. 
The inclusion of the KK modes relaxes the hierarchy among the model parameters 
that is required to realize a phenomenologically viable size of $r_{\rm c}$. 
We also discuss whether a nontrivial dilaton background enlarges $r_{\rm c}$ or not. 
\end{abstract}

\newpage

\section{Introduction}
The gravitational kinetic term is radiatively corrected by the quantum loop effect 
of the massive matters~\cite{Capper:1973bk,Adler:1980bx,Zee:1981mk}. 
Thus, when matter fields are localized on a 3-brane in a higher-dimensional gravitational theory, 
the four-dimensional (4D) Einstein-Hilbert term is induced on the brane. 
If the induced 4D Planck mass~$M_4$ is much larger than the higher dimensional one~$M_*$, 
the gravity that acts between two sources on the brane behaves like the 4D one 
for shorter distances than the crossover length scale~$r_{\rm c}$, which is defined by~\cite{Dvali:2000hr,Dubovsky:2002jm}
\be
 r_{\rm c} \equiv \begin{cases} \displaystyle \frac{M_4^2}{M_*^3} & (D=5) \\
 \displaystyle \frac{M_4}{M_*^2} & (D\geq 6)  \end{cases},  \label{r_c:vD}
\ee
where $D$ is the dimension of the spacetime, 
while behaves like the higher-dimensional one for longer distances than $r_{\rm c}$. 
This is called the DGP mechanism~\cite{Dvali:2000hr}. 
The crossover scale~$r_{\rm c}$ must be larger than the present Hubble size~$\sim 10^{26}$m 
from the consistency with the observational data. 

The DGP-type models have various interesting features when its cosmological time evolution 
is taken into account~\cite{Deffayet:2000uy}. 
It has the self-accelerating solution, in which the late-time accelerated expansion 
of the three-dimensional space occurs even in the absence of the cosmological constant. 
It also has the degravitating solution, which is relevant to the solution of the cosmological constant 
problem~\cite{Dvali:2002pe,ArkaniHamed:2002fu,Dvali:2007kt}. 

However, the five-dimensional (5D) DGP model is found to be disfavored by the observations~\cite{Xia:2009gb}. 
Thus we have to consider six- or higher- dimensional theories for phenomenologically viable models. 
When we work in such theories, the finite width of the brane has to be considered, 
otherwise the correlation functions diverge on the brane. 
So we consider a solitonic object in the field theory as the brane in this paper. 
The brane-localized modes are obtained as the low-lying Kaluza-Klein (KK) modes of a bulk field 
that couples to the soliton in such a case. 
Therefore, not only the low-lying KK modes but also higher-level KK modes that propagate into the bulk 
contribute to the induced gravity on the brane through the quantum loop effects. 

As can be seen from (\ref{r_c:vD}), we have to realize a huge hierarchy between the bulk gravitational scale~$M_*$ and 
the brane-induced Planck scale~$M_4$ in order to obtain a phenomenologically viable value of $r_{\rm c}$. 
Naively, this seems possible since we have an infinite number of the KK modes that contribute to $r_{\rm c}$. 
However, this is not so trivial because higher KK modes have larger KK masses and their couplings to 
the gravitational field are suppressed due to the rapid oscillating profiles in the extra dimensions 
so that their contributions to $r_{\rm c}$ are also suppressed. 

In this paper, we estimate the crossover scale~$r_{\rm c}$ for the brane-induced force 
when the brane is given by a solitonic field configuration, and discuss the possibility to realize a huge value of $r_{\rm c}$. 
Although the 5D DGP is observationally disfavored, it is instructive to understand the situation 
how the KK modes contribute to $r_{\rm c}$ in a 5D model. 
Hence we consider a 5D toy model with a domain-wall background in this paper. 
Besides, we neglect the tensor structure of the gravitational field and analyze a scalar field theory 
to simplify the situation. 

The paper is organized as follows. 
In the next section, we briefly review how the brane-induced force is generated 
through the quantum loop effect of a brane-localized mode. 
In Sec.~\ref{model}, we consider a scalar model that mimics the 5D gravitational theory 
with a domain wall background, perform the KK decomposition of the fields 
and derive explicit forms of the 5D propagators. 
In Sec.~\ref{CSscale}, we calculate the self-energy of the bulk field, 
expand the one-loop corrected propagator in terms of 4D momentum, 
and derive the expression of the crossover scale from 4D to 5D. 
In Sec.~\ref{dilaton:bck}, we discuss the effect of a nontrivial dilaton background, 
which was proposed to enlarge the crossover scale. 
Sec.~\ref{summary} is devoted to the summary. 
In Appendix~\ref{DWsector}, we complement the details of the domain-wall sector. 
In Appendix~\ref{Hyper_fct}, we collect the definitions and various properties of the special functions 
that are used in the text.

\section{Brane-induced force} \label{review}
Let us briefly review how the brane-induced force is generated by the quantum loop effect. 

We consider a 5D scalar theory with a brane located at $y=0$. 
The Lagrangian is given by~\footnote{
The indices~$M=0,1,2,3,4$ and $\mu=0,1,2,3$ are the 5D and 4D Lorentz indices, respectively. 
}
\be
 \cL = -\frac{M_5^3}{2}\der^M\Phi_{\rm G}\der_M\Phi_{\rm G}
 -\dlt(y)\brc{-\frac{1}{2}\der^\mu\phi\der_\mu\phi-\frac{m^2}{2}\phi^2-\lmd\Phi_{\rm G}\phi^2}, 
\ee
where the real scalars~$\Phi_{\rm G}(x^\mu,y)$ and $\phi(x^\mu)$ are a 5D bulk field and a 4D brane field, respectively. 
In this model, $\Phi_{\rm G}$ mimics the gravitational field whose tensor structure is neglected. 
The positive constant~$M_5$ is the 5D Planck mass. 
The 4D scalar~$\phi$ is a brane matter field whose ``gravitational'' interaction is parametrized by $\lmd$.\footnote{
The mass dimensions of the quantities are $[\Phi_{\rm G}]=0$, $[\phi]=1$, $[m]=1$ and $[\lmd]=2$, respectively.  
} 

The (tree-level) 5D propagator of $\Phi_{\rm G}$ is given by a solution of~\footnote{
We have moved to the momentum basis for the 4D coordinates~\cite{Giddings:2000mu,Gherghetta:2000kr}. 
} 
\be
 \brkt{-p^2+\der_y^2}G_{\rm G}(y,y';p) = -\frac{1}{M_5^3}\dlt(y-y'), \label{eq_for_GG}
\ee
where $p^2\equiv p^\mu p_\mu$, and has the following properties. 
\be
 G_{\rm G}(y,y';p) = G_{\rm G}(y',y;p), \;\;\;\;\;
 \lim_{y\to\pm\infty}\abs{G_{\rm G}(y,y';p)} < \infty.  \label{BD:GG}
\ee
We can easily solve the above equation, and obtain
\be
 G_{\rm G}(y,y';p) = \frac{1}{2M_5^3p}e^{-p\abs{y-y'}},  \label{trivial:prop}
\ee
where $p\equiv\sqrt{p^2}$. 
This tree-level propagator receives the quantum correction induced by the brane field~$\phi$. 
The quantum-corrected propagator~$\cG_{\rm G}(y,y';p)$ is given by
\bea
 \cG_{\rm G}(y,y';p) \eql G_{\rm G}(y,y';p)+G_{\rm G}(y,0;p)\Sgm_{\rm G}(p)G_{\rm G}(0,y';p) \nonumber\\
 &&+G_{\rm G}(y,0;p)\Sgm_{\rm G}(p)G_{\rm G}(0,0;p)\Sgm_{\rm G}(p)G_{\rm G}(0,y';p)+\cdots \nonumber\\
 \eql \frac{e^{-p\abs{y-y'}}-e^{-p(\abs{y}+\abs{y'})}}{2M_5^3p}
 +\frac{e^{-p(\abs{y}+\abs{y'})}}{2M_5^3p}\brc{1+\frac{\Sgm_{\rm G}(p)}{2M_5^3p}+\brkt{\frac{\Sgm_{\rm G}(p)}{2M_5^3p}}^2+\cdots} 
 \nonumber\\
 \eql \frac{e^{-p\abs{y-y'}}-e^{-p(\abs{y}+\abs{y'})}}{2M_5^3p}+\frac{e^{-p(\abs{y}+\abs{y'})}}{2M_5^3p-\Sgm_{\rm G}(p)}, 
 \label{cG_G:thin}
\eea
where $\Sgm_{\rm G}(p)$ is the self-energy of $\Phi_{\rm G}$. 
The one-loop contribution to the self-energy of $\Phi_{\rm G}$ is 
\be
 \Sgm_{\rm G}^{\rm 1loop}(p) = \lmd^2\int\frac{d^4q}{(2\pi)^4}\;\frac{1}{(q+p)^2+m^2}\frac{1}{q^2+m^2}. 
\ee
Here we expand $\Sgm_{\rm G}^{\rm 1loop}(p)$ in terms of the external momentum~$p$. 
\be
 \Sgm_{\rm G}^{\rm 1loop}(p) = \Sgm_{\rm G}^{(0)}+\Sgm_{\rm G}^{(1)}p+\Sgm_{\rm G}^{(2)}p^2+\cO(p^3). 
\ee
Then, we find that $\Sgm_{\rm G}^{(0)}$ is logarithmically divergent, $\Sgm_{\rm G}^{(1)}$ vanishes, and 
\be
 \Sgm_{\rm G}^{(2)} = -\frac{\lmd^2}{96\pi^2m^2}. 
\ee
Since $\Phi_{\rm G}$ mimics the gravitational field, we neglect $\Sgm_{\rm G}^{(0)}$. 
In the genuine gravity theory, it will vanish due to the invariance under the general coordinate transformation~\cite{Capper:1973bk}. 
Thus, the one-loop-corrected propagator between two sources located at the brane is calculated as
\bea
 \cG_{\rm G}^{\rm 1loop}(0,0;p) \eql \frac{1}{2M_5^3p-\Sgm_{\rm G}^{(2)}p^2+\cO(p^3)}
\eea
From this expression, we find that the ``gravitational force'' mediated by $\Phi_{\rm G}$ behaves like four-dimensional 
for shorter distances than the crossover scale~$r_{\rm c}$, which is defined by
\be
 r_{\rm c} \equiv \frac{|\Sgm_{\rm G}^{(2)}|}{2M_5^3} = \frac{\lmd^2}{192\pi^2M_5^3m^2}. \label{r_c:thin}
\ee
Note that the ``gravitational coupling''~$\lmd$ has mass-dimension~2. 

If all the mass scales are comparable, the crossover scale cannot be much larger than 
the compton length~$m^{-1}$ (or the 5D Planck length~$M_5^{-1}$). 
One simple way to enlarge $r_{\rm c}$ is to introduce 
vast number of matter fields on the brane~\cite{Dvali:2001gx}.  
If the model has $N$ copies of $\phi$ that equally interact with $\Phi_{\rm G}$, 
the crossover scale is enhanced by the factor~$N$. 
\be
 r_{\rm c} = \frac{N\lmd^2}{192\pi^2M_5^3m^2}. 
\ee
In fact, there are an infinite number of KK modes or stringy modes 
in compactified extra-dimensional models or the string theory. 
However, in these theories, higher modes have very large masses and thus 
their contribution to $r_{\rm c}$ is suppressed. 
Besides, the coupling of each mode to $\Phi_{\rm G}$ depends on the KK level, 
and those of higher modes are suppressed due to the rapidly oscillating profile in the extra dimension. 
Therefore, it is nontrivial whether $r_{\rm c}$ is finite or explicitly depends on the cutoff of the theory. 
In this paper, we will answer this question in a simple toy model.

\section{Thick-brane model} \label{model}
\subsection{Setup}
We consider the following 5D model. 
\bea
 \cL \eql -\frac{M_5^3}{2}\der^M\Phi_{\rm G}\der_M\Phi_{\rm G}
 -\frac{1}{2}\der^M\Phi_{\rm b}\der_M\Phi_{\rm b}-\frac{W(y)}{2}\Phi_{\rm b}^2
 -\lmd\Phi_{\rm G}\Phi_{\rm b}^2,  \label{cL}
\eea
where $\Phi_{\rm G}$ and $\Phi_{\rm b}$ are real scalar fields. 
As in the previous section, $\Phi_{\rm G}$ mimics the gravitational field whose tensor structure is neglected, 
and the low-lying KK modes of $\Phi_{\rm b}$ correspond to the brane-localized modes. 
Both of them are assumed to have vanishing backgrounds. 
The cubic coupling parametrized by $\lmd$ mimics the gravitational coupling. 

The real function~$W(y)$ represents the domain-wall background of some other scalar field. 
As a typical example, we assume that 
\be
 W(y) = A^2\tanh^2(By)+W_0,  \label{expr:W}
\ee
where $A$, $B$ and $W_0$ are real constants, and we take $A$ and $B$ to be positive 
(see Appendix~\ref{DWsector}). 
In this section, we neglect the warping of the spacetime induced by the domain wall. 
It will be considered in Sec.~\ref{warped}.

\subsection{``Gravitational field''~$\bdm{\Phi_{\rm G}}$}
The linearized equation of motion for $\Phi_{\rm G}$ is 
\be
 \brkt{\Box_4+\der_y^2}\Phi_{\rm G} = 0. 
\ee
Since the extra dimension is not compactified, the KK expansion of $\Phi_{\rm G}$ is 
\be
 \Phi_{\rm G}(x^\mu,y) = \int_0^\infty dm\;g_m(y)\chi_m(x^\mu), 
\ee
where the mode function~$g_m(y)$ is a solution of
\be
 \der_y^2g_m(y) = -m^2g_m(y). 
\ee
Namely, $g_m(y)$ is expressed by a linear combination of 
$\sin(my)$ and $\cos(my)$. 


The (tree-level) 5D propagator~$G_{\rm G}(y,y';p)$ satisfies the same equations 
as (\ref{eq_for_GG}) with (\ref{BD:GG}), and is given by (\ref{trivial:prop}).

\subsection{``Brane field''~$\bdm{\Phi_{\rm b}}$}
\subsubsection{KK expansion}
The linearized equation of motion for $\Phi_{\rm b}$ is 
\be
 \der^M\der_M\Phi_{\rm b}-W(y)\Phi_{\rm b} = 0. 
\ee
Thus, the KK expansion of $\Phi_{\rm b}$ is 
\be
 \Phi_{\rm b}(x^\mu,y) = \sumint{k} f_k(y)\phi_{k}(x^\mu), 
\ee
where the mode functions~$f_k(y)$ are solutions of the mode equations: 
\be
 \brc{\der_y^2-W(y))}f_k(y) = -m_k^2f_k(y),  \label{md_eq:phi}
\ee
with the KK masses~$m_k$. 

With (\ref{expr:W}), the mode equation~(\ref{md_eq:phi}) is expressed as
\be
 \brc{\der_y^2+m_k^2-A^2\tanh^2(By)-W_0}f_k(y) = 0.  \label{md_eq:phi:2}
\ee
Here we change the coordinate~$y$ as
\be
 y \to s \equiv \tanh(By). 
\ee
Then, the above equation becomes
\be
 \sbk{(1-s^2)\der_s^2-2s\der_s+a(a+1)-\frac{b^2}{1-s^2}}\tl{f}_k(s) = 0, \label{gLeq}
\ee
where $\tl{f}_k(s)\equiv f_k(y)$, and 
\be
 a \equiv -\frac{1}{2}+\sqrt{\frac{A^2}{B^2}+\frac{1}{4}}, \;\;\;\;\;
 b \equiv \frac{\sqrt{A^2+W_0-m_k^2}}{B}. 
\ee
This is the general Legendre equation, and thus its solution is expressed by 
a linear combination of the Legendre functions of the first and second kinds~$P_a^{b}(s)$ and $Q_a^{b}(s)$. 
To describe the mode functions, it is more convenient to choose 
\bea
 u_a^b(s) \defa (1-s^2)^{b/2}F\brkt{\frac{a+b+1}{2},\frac{-a+b}{2},\frac{1}{2};s^2}, \nonumber\\
 v_a^b(s) \defa (1-s^2)^{b/2}sF\brkt{\frac{a+b+2}{2},\frac{-a+b+1}{2},\frac{3}{2};s^2}, 
 \label{def:uv}
\eea
as two independent solutions of (\ref{gLeq}), 
where $F(\alp,\bt,\gm;z)={}_2F_1(\alp,\bt;\gm;z)$ is the hypergeometric function. 
The relations to $P_a^b(s)$ and $Q_a^b(s)$ are given by (\ref{rel:PQ-uv}) in Appendix~\ref{ano_exp:hyper}. 
Note that $u_a^b(s)$ and $v_a^b(s)$ are even and odd functions, respectively. 

As shown in Appendix~\ref{ano_exp:hyper}, 
the functions~$u_a^b(s)$ and $v_a^b(s)$ are even functions of the parameter~$b$, 
and thus can be rewritten as
\bea
 u_a^b(s) \eql (1-s^2)^{-b/2}F\brkt{\frac{a-b+1}{2},\frac{-a-b}{2},\frac{1}{2};s^2}, \label{expr:u:2} \\
 v_a^b(s) \eql (1-s^2)^{-b/2}sF\brkt{\frac{a-b+2}{2},\frac{-a-b+1}{2},\frac{3}{2};s^2}.  \label{expr:v:2}
\eea

We should also note that the mode functions~$f_k(y)$ should satisfy 
\be
 \lim_{y\to\pm\infty}\abs{f_k(y)} < \infty. \label{BC:inf}
\ee

\subsubsection{Localized modes}
Let us first consider a case of $m_k^2< A^2+W_0$. 
In this case, both $a$ and $b$ are positive. 
Thus, from (\ref{expr:u:2}) and (\ref{expr:v:2}), the necessary condition to satisfy (\ref{BC:inf}) is 
\be
 0 = F\brkt{\frac{a-b+1}{2},\frac{-a-b}{2},\frac{1}{2};1} = \frac{\sqrt{\pi}\Gm(b)}{\Gm(\frac{-a+b}{2})\Gm(\frac{a+b+1}{2})}, 
 \label{cond:F1}
\ee
or
\be
 0 = F\brkt{\frac{a-b+2}{2},\frac{-a-b+1}{2},\frac{3}{2};1} = \frac{\sqrt{\pi}\Gm(b)}{2\Gm(\frac{-a+b+1}{2})\Gm(\frac{a+b+2}{2})}. 
 \label{cond:F2}
\ee
We have used (\ref{F:BDv}) at the second equalities. 

The condition~(\ref{cond:F1}) indicates that $(-a+b)/2$ is zero or a negative integer. 
In this case, $u_a^b(s)$ in (\ref{def:uv}) can be rewritten by using (\ref{conv:u}) as
\be
 u_a^b(s) = \frac{\sqrt{\pi}\Gm(-b)}{\Gm(\frac{-a-b}{2})\Gm(\frac{a-b+1}{2})}(1-s^2)^{b/2}
 F\brkt{\frac{a+b+1}{2},\frac{-a+b}{2},b+1;1-s^2}. 
\ee
We can see that $u_a^b(\pm 1)$ vanishes.\footnote{
When $b$ is a non-negative integer, $\frac{a+b}{2}$ is also a non-negative integer. 
Thus, $\Gm(-b)/\Gm(\frac{-a-b}{2})$ remains finite.  
}

The condition~(\ref{cond:F2}) indicates that $(-a+b+1)/2$ is zero or a negative integer. 
In this case, $v_a^b(s)$ is rewritten by using (\ref{conv:v}) as
\be
 v_a^b(s) = \frac{\sqrt{\pi}\Gm(b)}{2\Gm(\frac{-a-b+1}{2})\Gm(\frac{a-b+2}{2})}(1-s^2)^{b/2}
 sF\brkt{\frac{a+b+2}{2},\frac{-a+b+1}{2},b+1;1-s^2}. 
\ee
Thus $v_a^b(\pm 1)$ vanishes. 

For example, in the case of (\ref{case:X}), the parameters become $a=2$ and $b=\sqrt{4-\hat{m}_k^2}$, 
where $\hat{m}_k\equiv m_k/B$. 
the allowed KK mass eigenvalues are
\be
 m_k^2 = k(4-k)B^2, 
\ee
where $k=0,1$.  
The corresponding mode functions are 
\bea
 f_0(y) \eql \cN_0u_2^2(\tanh(By)) = \frac{\cN_0}{\cosh^2(By)}, \nonumber\\
 f_1(y) \eql \cN_1v_2^1(\tanh(By)) = \frac{\cN_1\tanh(By)}{\cosh(By)},  \label{expr:f_01}
\eea
where $\cN_0=\sqrt{3B}/2$ and $\cN_1=\sqrt{3B}/\sqrt{2}$ are the normalization constants. 
These are localized modes around the domain wall.

\subsubsection{Bulk-propagating modes}
Next we consider a case of $m_k^2\geq A^2+W_0$. 
In this case, $b\equiv i\bt$ is pure imaginary. 
Since $\abs{(1-s^2)^{-b/2}}=\abs{e^{-\frac{i\bt}{2}\ln(1-s^2)}}=1$, we find that~\footnote{
Because $\Re b=0$ in this case, we cannot use the formula~(\ref{F:BDv}). 
}
\bea
 \lim_{s^2\nearrow 1}\abs{u_a^b(s)} \eql \abs{F\brkt{\frac{a-i\bt+1}{2},\frac{-a-i\bt}{2},\frac{1}{2};1}}, \nonumber\\
 \lim_{s^2\nearrow 1}\abs{v_a^b(s)} \eql \abs{F\brkt{\frac{a-i\bt+2}{2},\frac{-a-i\bt+1}{2},\frac{3}{2};1}}, 
\eea
which are finite. 
Thus the condition~(\ref{BC:inf}) is always satisfied. 
This indicates that there is a continuous spectrum above $m_k=\sqrt{A^2+W_0}$. 
The mode functions are expressed as~\footnote{
Note that when $u_a^b(s)$ and $v_a^b(s)$ are solutions of (\ref{gLeq}), 
so are $\brc{u_a^b(s)}^\dagger$ and $\brc{v_a^b(s)}^\dagger$. 
} 
\be
 f_k(y) = C_{1k}u_a^b(\tanh(By))+C_{2k}v_a^b(\tanh(By))+\hc, 
\ee
where $C_{1k}$ and $C_{2k}$ are constants.
Since $\lim_{y\to\pm\infty}\abs{f_k(y)}\neq 0$, these modes are not localized around the domain wall. 
In fact, using (\ref{conv:F}), we can rewrite the above solutions in the form of
\be
 \tl{f}_k(s) = \cC_{k+}(s)(1-s^2)^{b/2}+\cC_{k-}(s) (1-s^2)^{-b/2}+\hc, 
\ee
where $\cC_{k\pm}(s)$ depend on the parameters~$a$ and $b$, 
and take finite values at $s=\pm 1$. 
Since 
\bea
 (1-s^2)^{\pm b/2} \eql \exp\brc{\pm\frac{i\bt}{2}\ln(1-s^2)} 
 = \exp\brc{\mp i\bt\ln\cosh(By)} \nonumber\\
 \!\!\!&\sim \!\!\!& \exp\brc{\mp i\bt B{\rm sgn}\,(y)y}, 
\eea
for $\abs{y}\gg 1/B$, the above mode functions behave as plane-wave solutions at points far from the domain wall.

\subsubsection{5D propagator} \label{sec:G_b}
Here we derive the (tree-level) 5D propagator~$G_{\rm b}(y,y';p)$ for $\Phi_{\rm b}$ 
(see Appendix of Ref.~\cite{Gherghetta:2000kr} in the case of no domain wall background). 
It is defined as a solution of
\be
 \brc{-p^2+\der_y^2-W(y)}G_{\rm b}(y,y';p) = -\dlt(y-y'),  \label{eq:prop}
\ee
where $p^2=p^\mu p_\mu$, and $p_\mu$ is the Euclidean 4D momentum, 
and has the following properties. 
\be
 G_{\rm b}(y,y';p) = G_{\rm b}(y',y;p), \;\;\;\;\;
 \lim_{y\to\pm\infty}G_{\rm b}(y,y';p) = 0.  \label{G:prop}
\ee
In order to solve (\ref{eq:prop}), it is convenient to 
decompose $G_{\rm b}(y,y';p)$ into the following two parts. 
\be
 G_{\rm b}(y,y';p) = \vth(y-y')G_{{\rm b}>}(y,y';p)+\vth(y'-y)G_{\rm b<}(y,y';p). \label{decomp:G_b}
\ee
where $\vth(y)$ is the Heaviside step function, and 
\be
 \brc{-p^2+\der_y^2-W(y)}G_{\rm b\gtrless}(y,y';p) = 0.  \label{eq:G:hom}
\ee
Then, (\ref{G:prop}) is rewritten as
\be
 G_{\rm b>}(y,y';p) = G_{\rm b<}(y',y;p), \;\;\;\;\;
 \lim_{y\to\infty}G_{\rm b>}(y,y';p) = \lim_{y\to -\infty}G_{\rm b<}(y,y';p) = 0.  \label{G:prop:2}
\ee
Besides, by integrating (\ref{eq:prop}) for $y$ over the infinitesimal interval~$\sbk{y'-\ep,y'+\ep}$, 
we obtain the following matching condition. 
\be
 \left.\brc{\der_y G_{\rm b>}(y,y';p)-\der_y G_{\rm b<}(y,y';p)}\right|_{y'=y} = -1.  \label{G:match}
\ee

Note that (\ref{eq:G:hom}) is the same equation as (\ref{md_eq:phi}) if we replace $-p^2$ with $m_k^2$. 
Thus, its solution can be expressed by a linear combination of 
$u_a^{b(p)}(\tanh(By))$ and $v_a^{b(p)}(\tanh(By))$, where 
\be
 a \equiv -\frac{1}{2}+\sqrt{\frac{A^2}{B^2}+\frac{1}{4}}, \;\;\;\;\;
 b(p) \equiv \frac{\sqrt{A^2+W_0+p^2}}{B}.  \label{def:ab}
\ee
Taking (\ref{G:prop:2}) into account, $G_{\rm b\gtrless}(y,y';p)$ are expressed as
\bea
 G_{\rm b>}(y,y';p) \eql \cF_{a<}^{b(p)}(y')\cF_{a>}^{b(p)}(y), \nonumber\\
 G_{\rm b<}(y,y';p) \eql \cF_{a>}^{b(p)}(y')\cF_{a<}^{b(p)}(y), 
\eea
where $\cF_{a\gtrless}^b(s)$ is defined by~\footnote{
Since $u_a^b(s)$ and $v_a^b(s)$ are even and odd functions of $s$, 
we should note that $\cF_{a<}^b(y)=\cF_{a>}^b(-y)$. 
}
\bea
 \cF_{a\gtrless}^b(y) \defa \frac{R(a,b)}{\sqrt{2B}} u_a^b(\tanh(By)) 
 \mp \frac{v_a^b(\tanh(By))}{\sqrt{2B}R(a,b)},  \nonumber\\
 R(a,b) \defa \brc{\frac{\Gm(\frac{a+b+1}{2})\Gm(\frac{-a+b}{2})}
 {2\Gm(\frac{a+b+2}{2})\Gm(\frac{-a+b+1}{2})}}^{1/2}, 
 \label{def:cF}
\eea
so that they satisfy~\footnote{
See (\ref{behave:cF>:2}) in Appendix~\ref{ano_exp:hyper}. 
}
\be
 \lim_{y\to\infty}\cF_{a>}^{b(p)}(y) = \lim_{y\to -\infty}\cF_{a<}^{b(p)}(y) = 0. 
\ee
We can check that the condition~(\ref{G:match}) is satisfied by using the identity, 
\be
 u_a^{b}(s)\frac{dv_a^{b}}{ds}(s)-v_a^{b}(s)\frac{du_a^{b}}{ds}(s) = \frac{1}{1-s^2}. 
\ee
As a result, the 5D propagator is expressed as
\be
 G_{\rm b}(y,y';p) = \vth(y-y')\cF_{a<}^{b(p)}(y')\cF_{a>}^{b(p)}(y)+\vth(y'-y)\cF_{a>}^{b(p)}(y')\cF_{a<}^{b(p)}(y). 
 \label{expr:G_b}
\ee

From (\ref{md_eq:phi}) and (\ref{eq:prop}), the mode function~$f_k(y)$ is expressed as
\bea
 f_k(y) \eql \int_{-\infty}^\infty dy'\;\dlt(y-y')f_k(y') \nonumber\\
 \eql -\int_{-\infty}^\infty dy'\;\brc{-p^2+\der_{y'}^2-W(y')}G_{\rm b}(y,y';p)f_k(y') \nonumber\\
 \eql -\int_{-\infty}^\infty dy'\;G_{\rm b}(y,y';p)\brc{-p^2+\der_{y'}^2-W(y')}f_k(y') \nonumber\\
 \eql (p^2+m_k^2)\int_{-\infty}^\infty dy'\;G_{\rm b}(y,y';p)f_k(y'). 
\eea
We have performed the partial integral at the third equality. 
Thus, the 5D propagator is also expressed as
\be
 G_{\rm b}(y,y';p) = \sumint{k}\frac{f_k(y)f_k(y')}{p^2+m_k^2}. 
\ee

\section{Quantum correction to $\bdm{G_{\rm G}}$ and crossover scale} \label{CSscale}
\subsection{Quantum-corrected propagator}
The quantum-corrected 5D propagator~$\cG_{\rm G}(y,y';p)$ is given by
\bea
 \cG_{\rm G}(y,y';p) \eql G_{\rm G}(y,y';p)
 +\brkt{G_{\rm G}\cdot\Sgm_{\rm G}\cdot G_{\rm G}}(y,y';p) \nonumber\\
 &&+\brkt{G_{\rm G}\cdot\Sgm_{\rm G}\cdot G_{\rm G}\cdot\Sgm_{\rm G}\cdot G_{\rm G}}(y,y';p)+\cdots, 
 \label{full_prop}
\eea
where $\Sgm_{\rm G}(y,y';p)$ is the self-energy of $\Phi_{\rm G}$, 
and the dot denotes the integral over the extra-dimensional coordinate. 
For example, the second term is explicitly written as 
\be
 \brkt{G_{\rm G}\cdot\Sgm_{\rm G}\cdot G_{\rm G}}(y,y';p)
 \equiv \int_{-\infty}^\infty dy_1\int_{-\infty}^\infty dy_2\;
 G_{\rm G}(y,y_1;p)\Sgm_{\rm G}(y_1,y_2;p)G_{\rm G}(y_2,y';p).  \label{def:dot}
\ee

The 4D propagator between the sources on the brane is given by
\be
 G_{\rm 4D}(p) = \int_{-\infty}^\infty dy \int_{-\infty}^\infty dy'\; f_0(y)\cG_{\rm G}(y,y';p)f_0(y'). 
 \label{4Dpropagator}
\ee
where $f_0(y)$ denotes the mode function for the source localized on the domain-wall. 

The one-loop contribution to $\Sgm_{\rm G}$ is expressed as 
\be
 \Sgm_{\rm G}^{\rm 1loop}(y,y';p) = \int\frac{d^4q}{(2\pi)^4}\;\lmd^2
 G_{\rm b}(y,y';q+p)G_{\rm b}(y',y;q).  \label{expr:Sgm_G}
\ee
Since the integrand only depends on $q=\sqrt{q^2}$ and the angle~$\tht$ between the two vectors~$p^\mu$ and $q^\mu$ 
and is independent of the other two angles, the integrals for the latter angles can be trivially performed. 
Hence 
(\ref{expr:Sgm_G}) is rewritten as
\be
 \Sgm_{\rm G}^{\rm 1loop}(y,y';p) = \lmd^2\int_0^\infty\frac{dq}{4\pi^3}\;q^3
 \int_0^\pi  d\tht\;\sin^2\tht\;G_{\rm b}(y,y';q+p)G_{\rm b}(y';y;q). \label{expr:Sgm_G:2}
\ee

\subsection{4D momentum expansion of 5D propagator}
Here we expand the 4D propagator~$G_{\rm 4D}(p)$ 
in terms of the magnitude of 4D (Euclidean) momentum~$p=\sqrt{p^2}$. 

The tree-level 5D propagator~(\ref{trivial:prop}) is expanded as
\be
 G_{\rm G}(y,y';p) = \frac{1}{2M_5^3p}\brc{G_{\rm G}^{(0)}(y,y')+G_{\rm G}^{(1)}(y,y')p+\cO(p^2)}, 
\ee
where
\be
 G_{\rm G}^{(0)}(y,y') \equiv 1, \;\;\;\;\;
 G_{\rm G}^{(1)}(y,y') \equiv -\abs{y-y'}. \label{trivial:G0}
\ee

The self-energy~$\Sgm_{\rm G}^{\rm 1loop}$ in (\ref{expr:Sgm_G:2}) is expanded as
\bea
 \Sgm_{\rm G}^{\rm 1loop}(y,y';p) \eql \lmd^2\int_0^\infty \frac{dq}{4\pi^3}\;q^3\int_0^\pi d\tht\:\sin^2\tht\;
 \nonumber\\
 &&\hspace{5mm}
 \times\big[G_{\rm b}(y,y';q)+2q\der_{q^2}G_{\rm b}(y,y';q)\cdot p\cos\tht \nonumber\\
 &&\hspace{5mm}
 +\brc{\der_{q^2}G_{\rm b}(y,y';q)+2q^2\der_{q^2}^2G_{\rm b}(y,y';q)\cos^2\tht}p^2+\cO(p^3)\big]G_{\rm b}(y',y;q) \nonumber\\
 \eql \Sgm_{\rm G}^{(0)}(y,y')+\Sgm_{\rm G}^{(2)}(y,y')p^2+\cO(p^4), 
\eea
where the odd-power terms in $p$ vanish after the $\tht$-integration, and 
\bea
 \Sgm_{\rm G}^{(0)}(y,y') \defa \frac{\lmd^2}{8\pi^2}\int_0^\infty dq\;q^3G_{\rm b}^2(y,y';q), \nonumber\\
 \Sgm_{\rm G}^{(2)}(y,y') \defa \frac{\lmd^2}{16\pi^2}\int_0^\infty dq\;G_{\rm b}(y',y;q)
 \brc{2q^3\der_{q^2}G_{\rm b}(y,y';q)+q^5\der_{q^2}^2G_{\rm b}(y,y';q)}. 
 \label{expr:Sgm_G2}
\eea
As we did in Sec.~\ref{review}, we will neglect $\Sgm_{\rm G}^{(0)}$ in the following. 
In the genuine gravity theory, the $p$-independent term vanishes due to the invariance 
under the general coordinate transformation. 

Therefore, the 4D propagator~(\ref{4Dpropagator}) is expanded as
\bea
 G_{\rm 4D}(p) \eql f_0\cdot\brkt{G_{\rm G}+G_{\rm G}\cdot\Sgm_{\rm G}\cdot G_{\rm G}+\cdots}\cdot f_0 \nonumber\\
 \eql \frac{f_0\cdot G_{\rm G}^{(0)}\cdot f_0}{2M_5^3p}
 +f_0\cdot\brkt{\frac{G_{\rm G}^{(1)}}{2M_5^3}
 +\frac{G_{\rm G}^{(0)}\cdot\Sgm_{\rm G}^{(2)}\cdot G_{\rm G}^{(0)}}{4M_5^6}}\cdot f_0+\cO(p),  \label{G_4D}
\eea
where the dot denotes the $y$-integral like (\ref{def:dot}).

\subsection{Crossover scale}
The crossover scale~$r_{\rm c}$ is read off as the ratio of the expansion coefficients 
of the first two terms in (\ref{G_4D}). (See (\ref{cG_G:thin}) and (\ref{r_c:thin}).)
\be
 r_{\rm c} = \abs{\frac{1}{f_0\cdot G_{\rm G}^{(0)}\cdot f_0}\times
 f_0\cdot\brkt{G_{\rm G}^{(1)}+\frac{G_{\rm G}^{(0)}\cdot \Sgm_{\rm G}^{(2)}\cdot G_{\rm G}^{(0)}}{2M_5^3}}\cdot f_0}. 
 \label{expr:r_c}
\ee
Since $\Sgm_{\rm G}^{(2)}(y,y')$ is induced by a quantum correction induced by the $\Phi_{\rm b}$-loop, 
we are interested in the case that the second term in the parentheses dominates. 
Namely, $r_{\rm c}$ becomes 
\be
 r_{\rm c} \simeq \abs{\frac{f_0\cdot G_{\rm G}^{(0)}\cdot\Sgm_{\rm G}^{(2)}\cdot G_{\rm G}^{(0)}\cdot f_0}
 {2M_5^3\brkt{f_0\cdot G_{\rm G}^{(0)}\cdot f_0}}}.  \label{expr:r_c:2}
\ee

In the following, we assume that the zero-mode function~$f_0(y)$ is given by that in (\ref{expr:f_01}). 
Then, we can calculate 
\bea
 f_0\cdot G_{\rm G}^{(0)}\cdot f_0 \eql \int_{-\infty}^\infty dy\int_{-\infty}^\infty dy'\;
 f_0(y)G_{\rm G}^{(0)}(y,y')f_0(y') \nonumber\\
 \eql \brc{\int_{-\infty}^\infty dy\;f_0(y)}^2 
 = \frac{4\cN_0^2}{B^2}. 
%
\label{fGf}
\eea

From (\ref{expr:G_b}), the integrands in (\ref{expr:Sgm_G2}) are expressed by
\bea
 I_{\rm b}^{(1)}(y,y';q) \defa G_{\rm b}(y',y;q)\der_{q^2}G_{\rm b}(y,y';q) \nonumber\\
 \eql \vth(y-y')\brc{\cA_{>}^{(0)}(y)\cA_{<}^{(1)}(y')+\cA_{>}^{(1)}(y)\cA_{<}^{(0)}(y')} \nonumber\\
 &&+\vth(y'-y)\brc{\cA_{<}^{(0)}(y)\cA_{>}^{(1)}(y')+\cA_{<}^{(1)}(y)\cA_{>}^{(0)}(y')}, \nonumber\\
 I_{\rm b}^{(2)}(y,y';q) \defa G_{\rm b}(y',y;q)\der_{q^2}^2G_{\rm b}(y,y';q) \nonumber\\
 \eql \vth(y-y')\brc{\cA_{>}^{(0)}(y)\cA_{<}^{(2)}(y')
 +2\cA_{>}^{(1)}(y)\cA_{<}^{(1)}(y')+\cA_{>}^{(2)}(y)\cA_{<}^{(0)}(y')} \nonumber\\
 &&+\vth(y'-y)\brc{\cA_<^{(0)}(y)\cA_>^{(2)}(y')+2\cA_<^{(1)}(y)\cA_>^{(1)}(y')+\cA_<^{(2)}(y)\cA_>^{(0)}(y')}, \nonumber\\
 \label{cAcB:trivial}
\eea
where~\footnote{
Notice that all the $q$-dependence of $I_{\rm b}^{(1,2)}$ comes from $b(q)$, which is a function of $q^2$. 
}
\be
 \cA_{\gtrless}^{(n)}(y) \equiv \cF_{a\gtrless}^{b(q)}(y)\der_{q^2}^n\cF_{a\gtrless}^{b(q)}(y). \;\;\;\;\;
 (n=0,1,2)
\ee
Then, $\Sgm_{\rm G}^{(2)}(y,y')$ in (\ref{expr:Sgm_G2}) is expressed as
\bea
 \Sgm_{\rm G}^{(2)}(y,y') \eql \frac{\lmd^2}{16\pi^2}\int_0^\infty dq\;
 \brc{2q^3I_{\rm b}^{(1)}(y,y';q)+q^5I_{\rm b}^{(2)}(y,y';q)}. 
\eea
Thus, the numerator in (\ref{expr:r_c:2}) is 
\be
 f_0\cdot G_{\rm G}^{(0)}\cdot\Sgm_{\rm G}^{(2)}\cdot G_{\rm G}^{(0)}\cdot f_0
 = \frac{\lmd^2}{16\pi^2}\int_0^\infty dq\;\brc{f_0\cdot G_{\rm G}^{(0)}\cdot 
 \brkt{2q^3I_{\rm b}^{(1)}+q^5I_{\rm b}^{(2)}}\cdot G_{\rm G}^{(0)}\cdot f_0}, 
\ee
The integrand in the right-hand-side is calculated as
\bea
 \brkt{f_0\cdot G_{\rm G}^{(0)}\cdot I_{\rm b}^{(1)}\cdot G_{\rm G}^{(0)}\cdot f_0}(q)
 \eql \brc{\int_{-\infty}^\infty dy\;f_0(y)}^2\int_{-\infty}^\infty \int_{-\infty}^\infty dy'dy''\;I_{\rm b}^{(1)}(y',y'';q)
 \nonumber\\
 \eql \frac{8\cN_0^2}{B^2}\brc{J_{\rm b}^{(0,1)}(q)+J_{\rm b}^{(1,0)}(q)}, \nonumber\\
 \brkt{f_0\cdot G_{\rm G}^{(0)}\cdot I_{\rm b}^{(2)}\cdot G_{\rm G}^{(0)}\cdot f_0}(q)
 \eql \frac{8\cN_0^2}{B^2}\brc{J_{\rm b}^{(0,2)}(q)+2J_{\rm b}^{(1,1)}(q)+J_{\rm b}^{(2,0)}(q)}, 
 \label{fGSgmGf}
\eea
where
\bea
 J_{\rm b}^{(n,l)}(q) \defa \int_{-\infty}^\infty dy\;\cA_>^{(n)}(y)\int_{-\infty}^y dy'\;\cA_<^{(l)}(y'). 
 \label{def:J_b}
\eea
We have used that $\int_{-\infty}^\infty dy\int_y^\infty dy'=\int_{-\infty}^\infty dy'\int_{-\infty}^{y'}dy$. 
Since $\cA_<^{(n)}(y')=\cA_>^{(n)}(-y')$ and $\int_{-\infty}^\infty dy\int_{-\infty}^ydy' 
=\int_{-\infty}^\infty d(-y')\int_{-\infty}^{-y'} d(-y)$, we can see that
\be
 J_{\rm b}^{(n,l)}(q) = J_{\rm b}^{(l,n)}(q). 
\ee
Therefore, (\ref{fGSgmGf}) can be simplified as
\be
 f_0\cdot G_{\rm G}^{(0)}\cdot\Sgm_{\rm G}^{(2)}\cdot G_{\rm G}^{(0)}\cdot f_0 
 = \frac{\lmd^2\cN_0^2}{\pi^2B^2}\int_0^\infty dq\;
 \sbk{2q^3J_{\rm b}^{(0,1)}(q)+q^5\brc{J_{\rm b}^{(0,2)}(q)+J_{\rm b}^{(1,1)}(q)}}. 
\ee

As a result, the crossover scale is expressed as
\be
 r_{\rm c} \simeq \frac{\lmd^2}{8\pi^2M_5^3}\abs{\int_0^\infty dq\;
 \sbk{2q^3J_{\rm b}^{(0,1)}(q)+q^5\brc{J_{\rm b}^{(0,2)}(q)+J_{\rm b}^{(1,1)}(q)}}}. 
 \label{fin:expr:r_c}
\ee
In order to see whether this quantity is finite or not, the asymptotic behavior of the integrand for large $q$ is important. 
From (\ref{asymp:cA}), we find that  
\be
 \cA_>^{(n)}(y) \simeq \brkt{-\frac{y}{2q}}^n\frac{e^{-2qy}}{2q},  \label{asymp:cA>:flat}
\ee
for $\abs{q}\gg B$. 
This indicates that each $J_{\rm b}^{(n,l)}(q)$ diverges. 
This divergence comes from the fact that higher KK modes propagate in the bulk just like plane waves 
and the extra dimension has an infinite volume. 
However, a domain wall generically warps the ambient space, 
and its volume can be finite if we work in a gravitational theory. 
In the next subsection, we will see that $J_{\rm b}^{(n,l)}(q)$ become finite in the warped geometry.

\subsection{Warped geometry} \label{warped}
In the presence of the domain wall, 
\be
 \Phi_{\rm wall}(y) = v\tanh(By), 
\ee
the spacetime is warped with the metric
\be
 ds^2 = a_{\rm w}^2(y)\eta_{\mu\nu}dx^\mu dx^\nu+dy^2, 
\ee
where the warp factor~$a_{\rm w}(y)$ is given by~\cite{Kehagias:2000au,Shaposhnikov:2005hc} 
\be
 a_{\rm w}(y) = \cosh^{-2\bt}(By)\exp\brc{-\frac{\bt}{2}\tanh^2(By)}, \;\;\;\;\;
 \bt \equiv \frac{v^2}{9M_5^3}. 
\ee

We are interested in the asymptotic behaviors of the 5D propagator~$G_{\rm b}(y,y';q)$ for $q\gg B$ 
in a region far from the domain wall. 
In such a region, the above background geometry is approximated as the Randall-Sundrum spacetime~\cite{Randall:1999vf} 
with the warp factor
\be
 a_{\rm w}^{\rm RS}(y) = e^{-k\abs{y}}, \;\;\;\;\;
 k \equiv 2\bt B. 
\ee
We have rescaled the 4D coordinates~$x^\mu$ so that $a_{\rm w}^{\rm RS}(0)=1$.. 
Then, (\ref{eq:prop}) is modified as~\cite{Gherghetta:2000kr}
\be
 \brc{-e^{2k\abs{y}}p^2+\der_y^2-4k\,{\rm sgn}(y)\der_y-W_\infty}G_{\rm b}(y,y';p) = -e^{4k\abs{y}}\dlt(y-y'), 
 \label{eq:G_b:warped}
\ee
where $W_\infty \equiv A^2+W_0$. 
With the conditions in (\ref{G:prop}), this is solved as
\be
 G_{\rm b}(y,y';p) = \vth(y-y')\cF_<(y';p)\cF_>(y;p)+\vth(y'-y)\cF_>(y';p)\cF_<(y;p), 
\ee
where 
\bea
 \cF_>(y;p) \defa \begin{cases} \displaystyle \frac{e^{2ky}}{\sqrt{kH(\tl{p})}}K_\nu(\tl{p}e^{ky}) & (y>0) \\
 \displaystyle \sqrt{\frac{H(\tl{p})}{k}}e^{-2ky}\sbk{I_\nu(\tl{p}e^{-ky})
 -\frac{1}{K_\nu(\tl{p})}\brc{\frac{K_\nu(\tl{p})}{H(\tl{p})}+I_\nu(\tl{p})}K_\nu(\tl{p}e^{-ky})} & (y<0) \end{cases}, 
 \nonumber\\
 \cF_<(y;p) \defa \cF_>(-y;p), \nonumber\\
 H(\tl{p}) \defa -2K_\nu^2(\tl{p})-2\tl{p}K'_\nu(\tl{p})K_\nu(\tl{p}) \nonumber\\
 \eql -2K_\nu^2(\tl{p})+\tl{p}K_\nu(\tl{p})\brc{K_{\nu+1}(\tl{p})+K_{\nu-1}(\tl{p})}. 
\eea
Here, $\nu$ and $\tl{p}$ are defined by 
\bea
 \nu \defa \sqrt{4+\frac{W_\infty}{k}}, \;\;\;\;\;
 \tl{p} \equiv \frac{p}{k}, 
 \label{warped:cF}
\eea
and $I_\nu(z)$ and $K_\nu(z)$ are the modified Bessel functions of the first and second kinds, respectively. 

Now we focus on the case that $p\gg k$. 
Using (\ref{asymp:IK}), we find that $\cF_>(y;p)$ behaves as
\be
 \cF_>(y;p) \simeq \begin{cases} \displaystyle \frac{1}{\sqrt{2p}}\exp\brc{\tl{p}(1-e^{ky})+\frac{3}{2}ky} & (y>0) \\
 \displaystyle \frac{1}{\sqrt{2p}}\exp\brc{-\tl{p}(1-e^{-ky})-\frac{3}{2}ky} & (y<0) \end{cases}. 
 \label{approximate:cF>}
\ee
In the flat limit~$k\to 0$, this is reduced to
\be
 \cF_>(y;p) \simeq \frac{e^{-py}}{\sqrt{2p}}, 
\ee
which agrees with (\ref{asymp:cF>}) that leads to (\ref{asymp:cA>:flat}). 

Here we define the new coordinate~$Y$ as
\be
 Y \equiv {\rm sgn}(y)\frac{e^{k\abs{y}}-1}{k}. 
\ee
Then, (\ref{approximate:cF>}) is rewritten as
\be
 \tl{\cF}_>(Y;p) \equiv \cF_>(y;p) \simeq \frac{e^{-pY}}{\sqrt{2p}}(k\abs{Y}+1)^{3/2}. 
\ee
This leads to 
\bea
 \tl{\cA}_>^{(0)}(Y;q) \defa \tl{\cF}_>^2(Y;q) \simeq \frac{e^{-2qY}}{2q}(k\abs{Y}+1)^3, \nonumber\\
 \tl{\cA}_>^{(1)}(Y;q) \defa \tl{\cF}_>(Y;q)\der_{q^2}\tl{\cF}_>(Y;q) 
 \simeq \brkt{-\frac{Y}{2q}-\frac{1}{4q^2}}\frac{e^{-2qY}}{2q}(k\abs{Y}+1)^3, \nonumber\\
 \tl{\cA}_>^{(2)}(Y;q) \defa \tl{\cF}_>(Y;q)\der_{q^2}^2\tl{\cF}_>(Y;q)
 \simeq \brkt{\frac{Y^2}{4q^2}+\frac{Y}{2q^3}+\frac{5}{16q^4}}\frac{e^{-2qY}}{2q}(k\abs{Y}+1)^3. 
\eea
Using these quantities, $J_{\rm b}^{(n,l)}(q)$ defined in (\ref{def:J_b}) is now modified as
\bea
 J_{\rm b}^{(n,l)}(q) \eql \int_{-\infty}^\infty dy\;\sqrt{-G}\cA_>^{(n)}(y)
 \int_{-\infty}^y dy'\;\sqrt{-G}\cA_<^{(l)}(y') \nonumber\\
 \sma \int_{-\infty}^\infty \frac{dY}{(k\abs{Y}+1)^5}\;\tl{\cA}_>^{(n)}(Y;q)
 \int_{-\infty}^Y \frac{dY'}{(k\abs{Y'}+1)^5}\;\tl{\cA}_<^{(l)}(Y';q), 
\eea
where $\sqrt{-G}=a_{\rm w}^4(y)\simeq e^{-4k\abs{y}}=(k\abs{Y}+1)^{-4}$. 
Using (\ref{app:E_1}), these integrals are approximated as 
\bea
 J_{\rm b}^{(1,0)}(q) \eql J_{\rm b}^{(0,1)}(q) \simeq -\frac{1}{32kq^5}+\cO(\tl{q}^{-7}), \nonumber\\
 J_{\rm b}^{(2,0)}(q) \eql J_{\rm b}^{(0,2)}(q) \simeq \frac{1}{48k^3q^5}+\frac{19}{480kq^7}+\cO(\tl{q}^{-8}), \nonumber\\
 J_{\rm b}^{(1,1)}(q) \sma -\frac{1}{48k^3q^5}+\frac{1}{80kq^7}+\cO(\tl{q}^{-8}). 
\eea
Namely, we have
\be
 2q^3J_{\rm b}^{(0,1)}(q)+q^5\brc{J_{\rm b}^{(0,2)}(q)+J_{\rm b}^{(1,1)}(q)} 
 = -\frac{1}{96kq^2}+\cO(\tl{q}^{-3}). 
\ee
As a result, the integral in (\ref{fin:expr:r_c}) now converges. 
Since the approximate expression~(\ref{approximate:cF>}) is valid when $q\gg k$, 
and the integral for that region is dominant when $\bt$ in (\ref{warped:cF}) is small enough, 
the crossover scale is estimated as
\be
 r_{\rm c} \sim \frac{\lmd^2}{8\pi^2M_5^3}\int_k^\infty dq\;\frac{1}{96kq^2}
 = \frac{\lmd^2}{768\pi^2M_5^3k^2} = \frac{27M_5^3\lmd^2}{1024\pi^2v^4B^2}. 
\ee
In the flat geometry limit ($k\to 0$), this diverges as we saw in the previous subsection. 
However, when we take into account the warping effect of the geometry by the domain wall, 
we have a finite value of $r_{\rm c}$. 
Namely, we have to allow a hierarchy among the parameters
in order to obtain the crossover scale~$r_{\rm c}$ that is larger than the present Hubble radius~$\sim 10^{26}$~m, 
even after summing up contributions from an infinite number of KK modes. 
For example, in the case that the bulk gravitational scale is $M_5\sim\sqrt{\lmd}\sim 10^{11}$~GeV,  
the domain wall scale, which is characterized by $v^{2/3}$ and $B$, has to be less than $10^4$~GeV.

\section{Nontrivial dilaton background} \label{dilaton:bck}
In Refs.~\cite{Dvali:2001gm,Gabadadze:2007dv}, the authors proposed a mechanism to enlarge the crossover scale~$r_c$. 
The idea is to allow the coefficient of the kinetic term for $\Phi_{\rm G}$ in (\ref{cL}) 
to have a nontrivial $y$-dependence. 
\be
 \cL = -\frac{K(y)}{2}\der^M\Phi_{\rm G}\der_M\Phi_{\rm G}+\cdots, 
\ee
where $K(y)$ is a real function that satisfies 
\be
 \lim_{y\to\pm\infty}K(y) = M_5^3, \;\;\;\;\;
 K(0) \ll M_5^3. 
\ee
The function $K(y)$ is understood as a background field configuration of the dilaton. 

In Ref.~\cite{Gabadadze:2007dv}, the thin-wall limit (\ie, $B\to \infty$) is considered, and 
$K(y)$ is assumed to take a tiny (positive) value~$\ep M_5^3$ ($\ep\ll 1$) on the brane. 
After the canonical normalization of $\Phi_{\rm G}$, its coupling to $\Phi_{\rm b}$ is rescaled as $\lmd/\sqrt{K(y)}$. 
Since the quantum-induced 4D Planck mass is proportional to $\lmd/\sqrt{K(0)}$, 
the crossover scale~$r_c$ is enhanced by a factor~$1/\ep$. 

However, we have to be more careful to discuss this enhancement because they did not consider 
the internal structure of the brane and the profile of the ``gravitational field''~$\Phi_{\rm G}$ there. 
To illustrate the situation, we assume that
\be
 K(y) = M_5^3\tanh^2(B_Ky), 
\ee
where $B_K$ is a positive constant. 

Since the behavior of the propagator around the brane is essential in this mechanism, in this section, 
we neglect the warping of the spacetime, which mainly affects the behavior at positions far from the brane. 
On the nontrivial dilaton background, (\ref{eq_for_GG}) is modified as
\be
 \brkt{-p^2+\der_y^2-\frac{K''(y)}{2K(y)}+\frac{K^{\prime 2}(y)}{4K^2(y)}}\tl{G}_{\rm G}(y,y';p) = -\frac{1}{M_5^3}\dlt(y-y'), 
\ee
where $\tl{G}_{\rm G}(y,y';p)$ is the 5D propagator of the canonically normalized 
field~$\tl{\Phi}_{\rm G}\equiv \sqrt{K(y)}\Phi_{\rm G}$. 
We can solve this equation by the technique that we did in Sec.~\ref{sec:G_b}, 
\ignore{
Namely, 
\be
 \tl{G}_{\rm G}(y,y';p) = \vth(y-y')\tl{G}_{\rm G>}(y,y';p)+\vth(y-y')\tl{G}_{\rm G<}(y,y';p), 
\ee
where $\tl{G}_{\rm G\gtrless}(y,y';p)$ satisfy 
\be
 \brc{-p^2+\der_y^2-\frac{K''(y)}{2K(y)}+\frac{K^{\prime 2}(y)}{4K^2(y)}}\tl{G}_{\rm G\gtrless}(y,y';p) = 0, 
 \label{eq:tlG_G}
\ee
and
\bea
 &&\tl{G}_{\rm G>}(y,y';p) = \tl{G}_{\rm G<}(y',y;p), \nonumber\\
 &&\lim_{y\to\infty}\tl{G}_{\rm G>}(y,y';p) = \lim_{y\to -\infty}\tl{G}_{\rm G<}(y,y';p) = 0, \nonumber\\
 &&\left.\brc{\der_y\tl{G}_{\rm G>}(y,y';p)-\der_y\tl{G}_{\rm G<}(y,y';p)}\right|_{y'=y} = -\frac{1}{M_5^3}. 
\eea
Notice that (\ref{eq:tlG_G}) has the same form as (\ref{gLeq}) with $a\to 1$ and $b\to\hat{p}\equiv p/B_K$. 
}
and find that
\be
 \tl{G}_{\rm G}(y,y';p) = \vth(y-y')\cG_{<}^{\hat{p}}(y')\cG_{>}^{\hat{p}}(y)
 +\vth(y'-y)\cG_{>}^{\hat{p}}(y')\cG_{<}^{\hat{p}}(y), 
\ee
where $\hat{p}\equiv p/B_K$, and 
\be
 \cG_{\gtrless}^{\hat{p}}(y) \equiv \frac{R(1,\hat{p})}{\sqrt{2B_KM_5^3}}u_1^{\hat{p}}(\tanh(B_Ky))
 \mp\frac{v_1^{\hat{p}}(\tanh(B_Ky))}{\sqrt{2B_KM_5^3}R(1,\hat{p})}, 
\ee
which is similar to $\cF_{a\gtrless}^b(y)$ defined by (\ref{def:cF}). 

As shown in Appendix~\ref{ano_exp:hyper}, $u_1^{\hat{p}}(s)$ and $v_1^{\hat{p}}(s)$ are even functions of $\hat{p}$. 
Thus, they are expanded as
\bea
 u_1^{\hat{p}}(\tanh(B_Ky)) \eql (1-\tanh^2(B_Ky))^{\hat{p}/2}F\brkt{\frac{2+\hat{p}}{2},\frac{-1+\hat{p}}{2},\frac{1}{2};\tanh^2(B_Ky)} 
 \nonumber\\
 \eql 1-B_Ky\tanh(B_Ky)+\cO\brkt{\hat{p}^2}, \nonumber\\
 v_1^{\hat{p}}(\tanh(B_Ky)) \eql (1-\tanh^2(B_Ky))^{\hat{p}/2}\tanh(B_Ky)
 F\brkt{\frac{3+\hat{p}}{2},\frac{\hat{p}}{2},\frac{3}{2};\tanh^2(B_Ky)} \nonumber\\
 \eql \tanh(B_Ky)+\cO\brkt{\hat{p}^2}. 
\eea
Here we have used that 
\bea
 F\brkt{1,-\frac{1}{2},\frac{1}{2};\tanh^2(B_Ky)} \eql 1-B_Ky\tanh(B_Ky), \nonumber\\
 F\brkt{\frac{3}{2},0,\frac{3}{2};\tanh^2(B_Ky)} \eql 1. 
\eea 
Since 
\be
 R(1,\hat{p}) = \brc{\frac{\Gm(\frac{2+\hat{p}}{2})\Gm(\frac{-1+\hat{p}}{2})}{2\Gm(\frac{3+\hat{p}}{2})\Gm(\frac{\hat{p}}{2})}}^{1/2}
 = i\sqrt{\hat{p}}\brc{1+\cO(\hat{p}^2)}, 
\ee
the function~$\cG_{>}^{\hat{p}}(y)$ is expanded as
\bea
 \cG_{>}^{\hat{p}}(y) \eql \sbk{\frac{i\sqrt{\hat{p}}}{\sqrt{2B_KM_5^3}}\brc{1-B_Ky\tanh(B_Ky)}
 -\frac{\tanh(B_Ky)}{i\sqrt{2B_KM_5^3\hat{p}}}}\brc{1+\cO(\hat{p}^2)}. 
\eea
Thus, the 5D propagator~$\tl{G}_{\rm G}(y,y';p)$ is expanded as
\be
 \tl{G}_{\rm G}(y,y';p) = \frac{1}{2M_5^3p}\brc{\tl{G}_{\rm G}^{(0)}(y,y')+\tl{G}_{\rm G}^{(1)}(y,y')p+\cO(\hat{p}^2)}, 
 \label{expand:tlG_G}
\ee
where
\bea
 \tl{G}_{\rm G}^{(0)}(y,y') \defa \tanh(B_Ky')\tanh(B_Ky), \nonumber\\
 \tl{G}_{\rm G}^{(1)}(y,y') \defa -\frac{\abs{\tanh(B_Ky)-\tanh(B_Ky')}}{B_K}-\abs{y-y'}\tanh(B_Ky')\tanh(B_Ky).  \nonumber\\
 \label{G_G:dilaton}
\eea

The propagator of the ``brane field''~$\Phi_{\rm b}$, $G_{\rm b}(y,y';p)$, is unchanged from that in the previous section. 
Hence, the one-loop self-energy~$\tl{\Sgm}_{\rm G}(y,y';p)$ of $\tl{\Phi}_{\rm G}$ is now given by
\bea
 \tl{\Sgm}_{\rm G}(y,y';p) \eql \int\frac{d^4q}{(2\pi)^4}\;\frac{\lmd}{\sqrt{K(y)}}\frac{\lmd}{\sqrt{K(y')}}
 G_{\rm b}(y,y';q+p)G_{\rm b}(y',y;q) \nonumber\\
 \eql \tl{\Sgm}_{\rm G}^{(0)}(y,y')+\tl{\Sgm}_{\rm G}^{(2)}(y,y')p^2+\cO(p^4), 
\eea
where
\bea
 \tl{\Sgm}_{\rm G}^{(0)}(y,y') \defa \frac{\lmd^2}{8\pi^2\sqrt{K(y)K(y')}}
 \int_0^\infty dq\;q^3G_{\rm b}^2(y,y';q), \nonumber\\
 \tl{\Sgm}_{\rm G}^{(2)}(y,y') \defa \frac{\lmd^2}{16\pi^2\sqrt{K(y)K(y')}}
 \int_0^\infty dq\;G_{\rm b}(y',y;q)\brc{2q^3\der_{q^2}G_{\rm b}(y,y';q)
 +q^5\der_{q^2}^2G_{\rm b}(y,y';q)}. \nonumber\\  \label{Sgm_G:dilaton}
\eea
Note that these quantities are enhanced around the brane~$y=0$, as expected. 
Again, we drop the contribution of $\tl{\Sgm}_{\rm G}^{(0)}(y,y')$. 
Then, the crossover scale~$r_{\rm c}$ is obtained just in the same way as (\ref{expr:r_c:2}). 
\bea
 r_{\rm c} \sma \abs{\frac{f_0\cdot\tl{G}_{\rm G}^{(0)}\cdot\tl{\Sgm}_{\rm G}^{(2)}\cdot\tl{G}_{\rm G}^{(0)}\cdot f_0}
 {2M_5^3\brkt{f_0\cdot\tl{G}_{\rm G}^{(0)}\cdot f_0}}}. 
\eea
Using (\ref{G_G:dilaton}) and (\ref{Sgm_G:dilaton}), we have 
\bea
 f_0\cdot\tl{G}_{\rm G}^{(0)}\cdot f_0 \eql \brc{\int_{-\infty}^\infty dy\;f_0(y)\tanh(B_Ky)}^2, 
 \nonumber\\
 f_0\cdot\tl{G}_{\rm G}^{(0)}\cdot\tl{\Sgm}_{\rm G}^{(2)}\cdot\tl{G}_{\rm G}^{(0)}\cdot f_0
 \eql \brc{\int_{-\infty}^\infty dy\;f_0(y)\tanh(B_Ky)}^2 \nonumber\\
 &&\times\int_{-\infty}^\infty\int_{-\infty}^\infty dy_1dy_2\;
 \tanh(B_Ky_1)\tl{\Sgm}_{\rm G}^{(2)}(y_1,y_2)\tanh(B_Ky_2).  \nonumber\\
\eea
Note that the enhancement factor~$1/\sqrt{K(y)}$ contained in $\tl{\Sgm}_{\rm G}^{(2)}$ is exactly cancelled 
by $\tanh(B_Ky)$ coming from $\tl{G}_{\rm G}^{(0)}$. 
Therefore, the resultant expression of $r_{\rm c}$ is the same as (\ref{fin:expr:r_c}). 
Namely, this enhancement mechanism does not work. 

In fact, this result is independent of the detailed function form of $K(y)$. 
Since 
\be
 \brkt{\der_y^2-\frac{K''}{2K}+\frac{K^{\prime 2}}{4K^2}}\sqrt{K(y)} = 0, 
\ee
the leading term of the expansion~(\ref{expand:tlG_G}) has the form of 
\be
 \tl{G}_{\rm G}^{(0)}(y,y') = \frac{\sqrt{K(y)K(y')}}{M_5^3}. 
\ee
Thus, the enhancement factor~$1/\sqrt{K(y)}$ in $\tl{\Sgm}_{\rm G}^{(2)}(y,y')$ is always cancelled 
when we calculate the crossover scale~$r_{\rm c}$.

\section{Summary} \label{summary}
We have discussed the contribution of an infinite number of the KK modes 
to the brane-induced force when the brane is given by a domain wall in an uncompactified 5D theory. 
In particular, we estimated the crossover scale~$r_{\rm c}$ from 4D to 5D for the induced force, 
and clarified whether $r_{\rm c}$ is enhanced to be of the order of the Hubble radius 
by that contribution. 
Taking into account the warping of the fifth-dimensional space by the domain wall, 
$r_{\rm c}$ becomes finite. 
So we cannot realize a large value of $r_{\rm c}$ without introducing a hierarchy among 
the model parameters. 

The parameters in our model can be categorized into the ``gravitational'' ones~$\{\lmd,M_5\}$ 
and the ones relevant to the brane physics~$\{m,A,B,v,B_K\}$. 
If we assume that the parameters in each class are of the same order of magnitude, 
the crossover scale is roughly estimated as
\be
 r_{\rm c} \simeq \begin{cases} \displaystyle 10^{-3}\times\brkt{\frac{M_5}{m_{\rm b}}}^2\times l_5 & 
 \mbox{(single brane mode)} \\ 
 \displaystyle 10^{-3}\times\brkt{\frac{M_5}{m_{\rm b}}}^8\times l_5 & \mbox{(including KK modes)}
 \end{cases}, 
\ee
where $l_5\equiv 1/M_5$ is the 5D Planck length, and $m_{\rm b}$ denotes the mass scale of the brane physics. 
Thus, if we admit some hierarchy between $M_5$ and $m_{\rm b}$, 
the contribution of all the KK modes makes it easier to realize 
a phenomenologically viable size of $r_{\rm c}$. 

The authors of Refs.~\cite{Dvali:2001gm,Gabadadze:2007dv} proposed a mechanism to enlarge $r_{\rm c}$ 
by introducing a nontrivial dilaton background. 
After the canonical normalization of the ``gravitational field''~$\Phi_{\rm G}$, 
the ``gravitational couplings'' to the brane modes are enhanced in this case. 
However, this nontrivial background also affects the propagator of $\Phi_{\rm G}$, 
and in fact, this effect exactly cancels the enhancement of the above couplings. 
Hence this mechanism does not work in the 5D theory. 

In higher-dimensional theories, the situation may be different. 
Since we have more KK modes and less warping effects on the ambient geometry, 
we expect that $r_{\rm c}$ explicitly depends on the cutoff scale of the theory. 
Besides, the nontrivial dilaton background may enlarge $r_{\rm c}$ because 
the behavior of the propagator near the brane is quite different from the 5D case in such a higher-dimensional theory, 
and thus the cancellation of the enhancement factor will not occur.  
In order to clarify these points, we need to extend our study to higher dimensions. 
We will discuss these issues in separate papers.


\appendix

\section{Domain-wall sector} \label{DWsector}
Here we provide specific examples of the ``brane field''~$\Phi_{\rm b}$, 
and show their couplings have the form of (\ref{cL}) with (\ref{expr:W}). 

\subsection{Domain-wall background}
The simplest model that has a domain-wall background is 
\be
 \cL_X = -\frac{1}{2}\der^MX\der_MX+\frac{m_X^2}{2}X^2-\frac{\lmd_X}{4}X^4, \label{cL:X}
\ee
where $X$ is a real scalar field, and $m_X,\lmd_X>0$. 
The equation of motion is 
\be
 \der^M\der_M X+m_X^2 X-\lmd_X X^3 = 0. 
\ee
and thus, there are two degenerate vacua~$\vev{X}=\pm m_X/\sqrt{\lmd_X}$. 
When we take the boundary conditions~$\lim_{y\to\pm\infty}\vev{X}=\pm m_X/\sqrt{\lmd_X}$,  
this model has the domain wall solution,\footnote{
By shifting the coordinate~$y$, we can always set the position of the domain wall at $y=0$. 
} 
\be
 X_{\rm bg}(y) = \frac{m_X}{\sqrt{\lmd_X}}\tanh\brkt{\frac{m_X}{\sqrt{2}}y}.  \label{X_bg}
\ee

\subsection{Fluctuation modes around the domain wall}
Around this background, the scalar field~$X$ is divided as
\be
 X = X_{\rm bg}+\tl{X}, \label{X:flct}
\ee
where $\tl{X}$ is the fluctuation part. 
Plugging this into (\ref{cL:X}), we obtain 
\bea
 \cL_X \eql -\frac{1}{2}\der^M\tl{X}\der_M\tl{X}+\brkt{\frac{m_X^2}{2}-\frac{3\lmd_X}{2}X_{\rm bg}^2}\tl{X}^2
 -\lmd_X X_{\rm bg}\tl{X}^3-\frac{\lmd_X}{4}\tl{X}^4 \nonumber\\
 \eql -\frac{1}{2}\der^M\tl{X}\der_M\tl{X}
 -\frac{1}{2}\brc{-m_X^2+3m_X^2\tanh^2\brkt{\frac{m_X}{\sqrt{2}}y}}\tl{X}^2+\cO(\tl{X}^3), 
\eea
where we have dropped the $\tl{X}$-independent term and the total derivative term. 
Thus, at the quadratic order, the Lagrangian for $\tl{X}$ has the same form as $\Phi_{\rm b}$ in (\ref{cL}) 
if we identify $A$ and $B$ in (\ref{expr:W}) as 
\be
 A = \sqrt{3}m_X, \;\;\;\;\;
 B = \frac{m_X}{\sqrt{2}}, \;\;\;\;\;
 W_0 = -m_X^2.  \label{case:X}
\ee

\subsection{Matter field coupled to the domain wall}
Next we consider a matter field coupled to the domain-wall field~$X$ as
\be
 \cL_Q = -\frac{1}{2}\der^MQ\der_MQ-\frac{m_Q^2}{2}Q^2-\frac{\lmd_{XQ}}{2}Q^2X^2, 
\ee
where the matter field~$Q$ is a real scalar, and $m_Q,\lmd_{XQ}>0$. 
Plugging (\ref{X:flct}) into this, we obtain 
\bea
 \cL_Q \eql -\frac{1}{2}\der^MQ\der_MQ-\frac{1}{2}\brc{m_Q^2+\frac{\lmd_{XQ}}{\lmd_X}m_X^2\tanh^2\brkt{\frac{m_X}{\sqrt{2}}y}}Q^2, 
\eea
at the quadratic order in $Q$. 
This has the same form as (\ref{cL}) if we identify $\Phi_{\rm b}=Q$ and the constants in (\ref{expr:W}) as
\be
 A = \sqrt{\frac{\lmd_{XQ}}{\lmd_X}}m_X, \;\;\;\;\;
 B = \frac{m_X}{\sqrt{2}}, \;\;\;\;\;
 W_0 = m_Q^2.  \label{case:Q}
\ee

\section{Various properties of special functions} \label{Hyper_fct}
\subsection{General properties of hypergeometric functions}
The hypergeometric function~$F(\alp,\bt,\gm;z)={}_2F_1(\alp,\bt;\gm;z)$ is defined by
\bea
 F(\alp,\bt,\gm;z) \eql \frac{\Gm(\gm)}{\Gm(\alp)\Gm(\bt)}\sum_{n=0}^\infty\frac{\Gm(\alp+n)\Gm(\bt+n)}{\Gm(\gm+n)}\frac{z^n}{n!} 
 \nonumber\\
 \eql 1+\frac{\alp\bt}{\gm}z+\frac{\alp(\alp+1)\bt(\bt+1)}{2\gm(\gm+1)}z^2+\cO(z^3),  \label{def:hyper_fct}
\eea
where $\Gm(z)$ is the gamma function, and satisfies
\be
 \sbk{z(1-z)\der_z^2+\brc{\gm-(\alp+\bt+1)z}\der_z-\alp\bt}F(\alp,\bt,\gm;z) = 0. 
\ee
This function can be rewritten as
\bea
 F(\alp,\bt,\gm;z) \eql \frac{\Gm(\gm)\Gm(\alp+\bt-\gm)}{\Gm(\alp)\Gm(\bt)}(1-z)^{\gm-\alp-\bt}
 F(\gm-\alp,\gm-\bt,\gm-\alp-\bt+1;1-z) \nonumber\\
 &&+\frac{\Gm(\gm)\Gm(\gm-\alp-\bt)}{\Gm(\gm-\alp)\Gm(\gm-\bt)}
 F(\alp,\bt,\alp+\bt-\gm+1;1-z).  \label{conv:F}
\eea
We also have the formula:
\be
 F(\alp,\bt,\gm;1) = \frac{\Gm(\gm)\Gm(\gm-\alp-\bt)}{\Gm(\gm-\alp)\Gm(\gm-\bt)}, 
 \label{F:BDv}
\ee
for $\Re\gm>0$ and $\Re(\gm-\alp-\bt)>0$. 

The derivative of $F(\alp,\bt,\gm;z)$ in terms of $z$ is given by
\be
 \der_z F(\alp,\bt,\gm;z) = \frac{\alp\bt}{\gm}F(\alp+1,\bt+1,\gm+1;z). 
 \label{deriv:F}
\ee

\subsection{Gamma function}
The gamma function~$\Gm(z)$ satisfies that
\bea
 &&\Gm(\alp)\Gm(1-\alp) = \frac{\pi}{\sin(\pi\alp)}, \nonumber\\
 &&\Gm(2z) = \frac{2^{2z-1}}{\sqrt{\pi}}\Gm(z)\Gm\brkt{z+\frac{1}{2}}, \nonumber\\
%
%
 &&\lim_{z\to \infty}\frac{\Gm(z+\alp)}{z^\alp\Gm(z)} = 1. 
 \label{formula:Gamma}
\eea

\ignore{
The following integral is also useful. 
\be
 \int_{-\infty}^\infty \;\frac{dz}{\cosh^a z} = \frac{\sqrt{\pi}\Gm(\frac{a}{2})}{\Gm(\frac{a+1}{2})}. 
 \label{int:cosh}
\ee
}

\subsection{Modified Bessel functions}
The modified Bessel functions~$I_\nu(z)$ and $K_\nu(z)$ satisfy the relation, 
\be
 I_\nu(z)K'_\nu(z)-I'_\nu(z)K_\nu(z) = -\frac{1}{z}.  \label{Wronskian_rel}
\ee
For $z\gg 1$, they are approximated as
\be
 I_\nu(z) \simeq \frac{e^z}{\sqrt{2\pi z}}, \;\;\;\;\;
 K_\nu(z) \simeq \sqrt{\frac{\pi}{2z}}e^{-z}.  \label{asymp:IK}
\ee

\subsection{Exponential integral}
The exponential integral~${\rm E}_1(x)$ is defined by 
\be
 {\rm E}_1(x) \equiv \int_x^\infty dt\;\frac{e^{-t}}{t}. 
\ee
For $\abs{x}\gg 1$, this can be approximated by
\be
 {\rm E}_1(x) = \frac{e^{-x}}{x}\brc{1-\frac{1!}{x}+\frac{2!}{x^2}+\cO(3!x^{-3})}. 
 \label{app:E_1}
\ee

For example, the following integral is expressed by ${\rm E}_1(x)$. 
\bea
 \int_{-\infty}^Y dY'\;\frac{e^{2qY}}{2q(-kY+1)^2}
 \eql \sbk{\frac{e^{2pY'}}{2qk(1-kY')}-\frac{e^{2\hat{q}}}{k^2}{\rm E}_1(2\hat{q}(1-kY'))}_{-\infty}^Y \nonumber\\
 \eql \frac{e^{2qY}}{2qk(1-kY)}-\frac{e^{2\hat{q}}}{k^2}{\rm E}_1(2\hat{q}(1-kY)) \nonumber\\
 \eql \frac{e^{2qY}}{4q^2(1-kY)^2}-\frac{ke^{2qY}}{4q^3(1-kY)^3}
 +\frac{3k^2e^{2qY}}{8q^4(1-kY)^4}+\cO(\hat{q}^{-5}). \nonumber\\
\eea
We have used (\ref{app:E_1}) at the last equality.

\subsection{Properties of the mode functions} \label{ano_exp:hyper}
The Legendre functions~$P_a^b(s)$ and $Q_a^b(s)$ are expressed 
in terms of $u_a^b(s)$ and $v_a^b(s)$ in (\ref{def:uv}) as
\bea
 P_a^{b}(s) \eql \frac{2^b}{\sqrt{\pi}}\brc{\cos\brkt{\frac{a+b}{2}\pi}\hat{u}_a^b(s)+\sin\brkt{\frac{a+b}{2}\pi}\hat{v}_a^b(s)}, 
 \nonumber\\
 Q_a^{b}(s) \eql 2^{b-1}\sqrt{\pi}\brc{-\sin\brkt{\frac{a+b}{2}\pi}\hat{u}_a^b(s)
 +\cos\brkt{\frac{a+b}{2}\pi}\hat{v}_a^b(s)},  \label{rel:PQ-uv}
\eea
where
\be
 \hat{u}_a^b(s) \equiv \frac{\Gm(\frac{a+b+1}{2})}{\Gm(\frac{a-b+2}{2})}u_a^b(s), \;\;\;\;\;
 \hat{v}_a^b(s) \equiv \frac{2\Gm(\frac{a+b+2}{2})}{\Gm(\frac{a-b+1}{2})}v_a^b(s). 
\ee

We can show that $u_a^b(s)$ and $v_a^b(s)$ are 
even functions of the parameter~$b$. 
By using (\ref{conv:F}), we find that
\bea
 &&F\brkt{\frac{a+b+1}{2},\frac{-a+b}{2},\frac{1}{2};s^2} \nonumber\\
 \eql \frac{\sqrt{\pi}\Gm(b)}{\Gm(\frac{a+b+1}{2})\Gm(\frac{-a+b}{2})}(1-s^2)^{-b}
 F\brkt{\frac{-a-b}{2},\frac{a-b+1}{2},-b+1;1-s^2} \nonumber\\
 &&+\frac{\sqrt{\pi}\Gm(-b)}{\Gm(\frac{-a-b}{2})\Gm(\frac{a-b+1}{2})}
 F\brkt{\frac{a+b+1}{2},\frac{-a+b}{2},b+1;1-s^2},  \label{conv:u}
\eea
and
\bea
 &&F\brkt{\frac{a+b+2}{2},\frac{-a+b+1}{2},\frac{3}{2};s^2} \nonumber\\
 \eql \frac{\sqrt{\pi}\Gm(b)}{2\Gm(\frac{a+b+2}{2})\Gm(\frac{-a+b+1}{2})}(1-s^2)^{-b}
 F\brkt{\frac{-a-b+1}{2},\frac{a-b+2}{2},-b+1;1-s^2} \nonumber\\
 &&+\frac{\sqrt{\pi}\Gm(-b)}{2\Gm(\frac{-a-b+1}{2})\Gm(\frac{a-b+2}{2})}
 F\brkt{\frac{a+b+2}{2},\frac{-a+b+1}{2},b+1;1-s^2}. \label{conv:v}
\eea
Thus, we have
\bea
 u_a^b(s) \eql \cU_a^b(s^2)+\cU_a^{-b}(s^2), \nonumber\\
 v_a^b(s) \eql s\brc{\cV_a^b(s^2)+\cV_a^{-b}(s^2)},  \label{expr:uv:another}
\eea
where
\bea
 \cU_a^b(z) \defa \frac{\sqrt{\pi}\Gm(-b)}{\Gm(\frac{-a-b}{2})\Gm(\frac{a-b+1}{2})}
 (1-z)^{b/2}F\brkt{\frac{a+b+1}{2},\frac{-a+b}{2},b+1;1-z} \nonumber\\
 \eql C_u(b)(1-z)^{b/2}H_1(a,b;1-z), \nonumber\\
 \cV_a^b(z) \defa \frac{\sqrt{\pi}\Gm(-b)}{2\Gm(\frac{-a-b+1}{2})\Gm(\frac{a-b+2}{2})}
 (1-z)^{b/2}F\brkt{\frac{a+b+2}{2},\frac{-a+b+1}{2},b+1;1-z} \nonumber\\
 \eql C_v(b)(1-z)^{b/2}H_2(a,b;1-z). 
 \label{def:cUcV}
\eea
Here $H_1(a,b;w)\equiv F\brkt{\frac{a+b+1}{2},\frac{-a+b}{2},b+1;w}$ 
and $H_2(a,b;w)\equiv F\brkt{\frac{a+b+2}{2},\frac{-a+b+1}{2},b+1;w}$. 
The functions~$C_u(b)$ and $C_v(b)$ are defined as
\be
 C_u(b) = \frac{\sqrt{\pi}\Gm(-b)}{\Gm(\frac{-a-b}{2})\Gm(\frac{a-b+1}{2})}, \;\;\;\;\;
 C_v(b) = \frac{\sqrt{\pi}\Gm(-b)}{2\Gm(\frac{-a-b+1}{2})\Gm(\frac{a-b+2}{2})}. 
\ee
From the expression~(\ref{expr:uv:another}), we can see that $u_a^b(s)$ and $v_a^b(s)$ are 
even functions of $b$. 

Next we see the behaviors of the functions~$\cF_{a\gtrless}^b(y)$ defined in (\ref{def:cF}) 
near infinity~$y=\pm\infty$. 
We should note that
\be
 R(a,b) = \sqrt{\frac{C_v(-b)}{C_u(-b)}}. 
\ee
Thus, for $y\geq 0$ ($s\geq 0$), $\cF_{a>}^b(y)$ is expressed as
\bea
 \cF_{a>}^b(y) \eql \frac{R(a,b)}{\sqrt{2B}}u_a^b(s)-\frac{v_a^b(s)}{\sqrt{2B}R(a,b)} \nonumber\\
 \eql \frac{1}{\sqrt{2BC_u(-b)C_v(-b)}}\brc{C_v(-b)u_a^b(s)-C_u(-b)v_a^b(s)} \nonumber\\
 \eql \frac{(1-s^2)^{b/2}}{2b\sqrt{2BC_u(-b)C_v(-b)}}F\brkt{\frac{a+b+1}{2},\frac{-a+b}{2},b+1;1-s^2}, 
 \label{behave:cF>}
\eea
where $s=\tanh(By)$.  
At the last equality, we have used (\ref{def:uv}) and 
\bea
 &&F\brkt{\frac{a+b+1}{2},\frac{-a+b}{2},b+1;1-s^2} \nonumber\\
 \eql \frac{\Gm(b+1)\Gm(-\frac{1}{2})}{\Gm(\frac{a+b+1}{2})\Gm(\frac{-a+b}{2})}
 \abs{s}F\brkt{\frac{-a+b+1}{2},\frac{a+b+2}{2},\frac{3}{2};s^2} \nonumber\\
 &&+\frac{\Gm(b+1)\Gm(\frac{1}{2})}{\Gm(\frac{-a+b+1}{2})\Gm(\frac{a+b+2}{2})}
 F\brkt{\frac{a+b+1}{2},\frac{-a+b}{2},\frac{1}{2};s^2},  \nonumber\\
\eea
which follows from (\ref{conv:F}). 
Since $F(\alp,\bt,\gm;0)=1$, (\ref{behave:cF>}) indicates that
\bea
 \cF_{a>}^b(y) \sma \frac{(1-s^2)^{b/2}}{2b\sqrt{2BC_u(-b)C_v(-b)}} 
 = \frac{\cosh^{-b}(By)}{2b\sqrt{2BC_u(-b)C_v(-b)}} \nonumber\\
 \sma \frac{2^{b-1}e^{-bBy}}{b\sqrt{2BC_u(-b)C_v(-b)}}, 
 \label{behave:cF>:2}
\eea
for $y\gg 1/B$. 

For $y\ll -1/B$ (\ie, $s\simeq -1$), we find that
\bea
 \cF_{a>}^b(y) \sma \sqrt{\frac{2C_u(-b)C_v(-b)}{B}}(1-s^2)^{-b/2} 
 = \sqrt{\frac{2C_u(-b)C_v(-b)}{B}}\cosh^b(By) \nonumber\\
 \sma \frac{1}{2^b}\sqrt{\frac{2C_u(-b)C_v(-b)}{B}}e^{-bBy}. 
\eea
We have used (\ref{expr:uv:another}) and that $H_1(a,-b,1-s^2)\simeq H_2(a,-b,1-s^2)\simeq 1$. 

Using the second relation in (\ref{formula:Gamma}), we have 
\bea
 C_u(-b)C_v(-b) \eql \frac{\pi\Gm^2(b)}{2\Gm(\frac{-a+b}{2})\Gm(\frac{a+b+1}{2})\Gm(\frac{-a+b+1}{2})\Gm(\frac{a+b+2}{2})} 
 \nonumber\\
 \eql \frac{2^{2b-2}\Gm^2(b)}{\Gm(-a+b)\Gm(a+b+1)}
\eea
Therefore, the asymptotic behaviors of $\cF_{a>}^b(y)$ are 
\be
 \cF_{a>}^b(y) \simeq \begin{cases} \displaystyle \frac{\sqrt{\Gm(-a+b)\Gm(a+b+1)}}{\sqrt{2B}b\Gm(b)}e^{-bBy} & 
 \displaystyle \brkt{y \gg \frac{1}{B}} \\
 \displaystyle \frac{\Gm(b)}{\sqrt{2B\Gm(-a+b)\Gm(a+b+1)}}e^{-bBy} & 
 \displaystyle \brkt{y\ll -\frac{1}{B}} \end{cases}. \label{asymp:cF}
\ee
We can obtain the asymptotic behaviors of $\cF_{a<}^b(y)$ by using the relation~$\cF_{a<}^b(y)=\cF_{a>}^b(-y)$. 

When the loop momentum~$q$ is large enough, the parameter~$b(q)$ defined in (\ref{def:ab}) is approximated as
\be
 b(q) \simeq \frac{q}{B} \equiv \hat{q} \gg 1. 
\ee
Thus, since 
\be
 \Gm(b+c) \simeq b^c\Gm(b), 
\ee
for $b\gg c$, (\ref{asymp:cF}) is simplified as 
\be
 \cF_{a>}^{b(q)}(y) \simeq \frac{e^{-qy}}{\sqrt{2B\hat{q}}} = \frac{e^{-qy}}{\sqrt{2q}}, 
 \label{asymp:cF>}
\ee
for $\abs{y}\gg 1/B$. 
Therefore, we have
\be
 \cA_{>}^{(0)}(y) \simeq \frac{e^{-2qy}}{2q}, \;\;\;\;\;
 \cA_>^{(1)}(y) \simeq -\frac{y}{4q^2}e^{-2qy}, \;\;\;\;\;
 \cA_>^{(2)}(y) \simeq \frac{y^2}{8q^3}e^{-2qy}.  \label{asymp:cA}
\ee


\end{document}